\documentclass[11pt,a4paper]{article}
\usepackage[T1]{fontenc}
\usepackage[utf8]{inputenc}
\usepackage{lmodern}
\usepackage{microtype}
\usepackage[american]{babel}
\usepackage{csquotes}
\usepackage[a4paper,width=14cm,top=3cm,bottom=3.5cm]{geometry}

\usepackage{fancyhdr}
\usepackage{hyphenat}
\usepackage[backend=biber,style=apa,sorting=nyt]{biblatex}
\DeclareFieldFormat{doi}{DOI: \nolinkurl{#1}}
\usepackage{hyperref}
\title{\textbf{Statistical Leadership of What? \\Statistics After AI}}
\author{Anders Gorst-Rasmussen \\ {\small Novo Nordisk A/S, Søborg, Denmark} \\  \small \href{mailto:agtr@novonordisk.com}{\texttt{agtr@novonordisk.com}}}
\date{August 30, 2026}

\begin{document}
\maketitle

\begin{abstract}
\noindent
Statisticians have spent over a century arguing that we are more than calculators, usually by pointing to what else we know. AI is making that defense harder, since the list of what only statisticians can do grows shorter with each model release. AI makes claims cheap to generate and may eventually make the statistics behind them cheap too. However, a model cannot be answerable in the way that statistical practice requires. Statistical leadership then becomes a question of which claims we are there to answer for, including the ones we answer for in advance by building judgment into systems.
\end{abstract}

\noindent%
{\it Keywords:}  role of the statistician; answerability; artificial intelligence; automation; clinical trial; drug development
\section{A Very Able Calculator}
\label{sec:intro}

My first job as a statistician was consulting at a Danish hospital. The physician who hired me officially registered me as a \enquote{statistical calculator} (\emph{statistisk beregner}). The year was 2004, so I am fairly sure it was a joke. As a newly trained mathematical statistician, I did not mind. Most statisticians have met some version of this---usually unstated and not always in fun.
 
Our profession has been debating its own role for a long time. In 1849, the London Statistical Society noted in its annual report that one can  be \enquote{a very able calculator and yet but a second-rate statist} \parencite{statisticalsocietyFifteenthAnnualReport1849}. \textcite{brossRoleStatisticianScientist1974} asked whether the practicing statistician should think of himself as a scientist or a shoe clerk. Thirteen years later, \textcite{marquardtImportanceStatisticians1987} answered that statisticians must become entrepreneurs. 

Others have been picking up the calculating. \textcite{hahnKeyChallengesStatisticians1998} observed that we increasingly see \enquote{statistics without statisticians}. Over the last 25 years, statistical software escaped statistician supervision, data proliferated, and data science emerged as something new, professionally at least \parencite{donoho50YearsData2017}. And now AI. Where earlier shifts lowered the cost of running an analysis someone had decided to do, this one lowers the cost of deciding \emph{what} to do. 

What worries me is a seemingly mundane scenario. It is Tuesday afternoon. A colleague brings me a finding. They fed an AI model a dataset and a question, and it returned what looks like a complete analysis with hedges in all the right places. They want to know: \emph{can they say it?} I did not scope the analysis and I have no idea how many data cuts were tried. Now multiply that by everyone else who can do the same.

What do I take on? How do I take it on? Those are leadership questions, but not the kind statistical leadership advice has a clear answer to. It is more about conduct---how to get into the room, how to communicate, how to influence---than about \emph{what} we are leading \parencite{gibsonLeadershipStatisticsIncreasing2019}. Perhaps because that is supposed to be obvious.

So what exactly are statisticians leading? I will work with a definition sharper than \enquote{statistical thinking} and less self-flattering than \enquote{scientific rigor}: statistical practice means bringing discipline to how data can warrant claims, and statistical leadership means deciding where that discipline is applied and answering for the way it is applied. In other words, statistical leadership is more about responsibilities than capabilities or conduct. 

As claims become cheap to generate, statistical leadership is what statisticians are for. That is neither new nor especially reassuring, since this kind of leadership is not ours by default nor ours alone. I make this argument in a pharmaceutical setting; answerability is more explicit there than in most places, which makes it easier to see both what it consists of and what happens when it goes missing.

\section{What Are Statisticians For?}
\label{sec:what-for}

How does one \enquote{bring discipline to how data can warrant claims}? That depends on the claim:
\begin{itemize}
  \item In a registrational clinical trial, the claim is what the label will say, and the data have not yet been collected. The discipline is the estimand: deciding which treatment effect we mean determines what data must be collected \parencite{ichE9R1Estimands2019}.
 \item For a prediction model, the claim is how the method performs beyond its training data. The discipline is fixing the performance measure and validation set before seeing the result.
  \item In exploratory work, the claim is that something is worth looking into further. The discipline is transparency about how the conclusion arose---how many cuts of the data were tried, which thresholds were considered---and what other conclusions might be plausible.
\end{itemize}
So discipline can be procedure, mathematics, or plain honesty about what you did. None of this strictly requires a statistician. Indeed, the ASA defines participation in statistical practice by activity rather than job title \parencite{asaEthicalGuidelines2022}. What makes statistics a profession is not the activities but their aim. A clinician is primarily concerned with the clinical claim, and a data scientist with a model that generalizes. I will use \enquote{statistician} for someone whose professional focus is the warrant connecting data to the claim, whatever the title on their contract says. 

A warrant is rarely obvious, so it depends in part on trust, especially when the claim lands with someone not involved in producing it. In a small research team, trust is personal. In my consulting days, if I told my clinician peers that a particular kind of analysis was needed, they usually took my word for it. This changes when a claim has to travel to a regulator or a journal referee who has not met you, cannot rerun your study, and assumes you have an interest in the result. Procedure is what travels instead. We are trained to think of pre-specification mathematically, as multiplicity control, but it is also a way to state a judgment in a form others can check. \textcite{porterTrustNumbersPursuit1995} called this mechanical objectivity: institutions rely on impersonal procedure when personal standing is unavailable or compromised by interest. 

Drug development runs on mechanical objectivity because the conflict of interest is structural and claims end up on the drug label. Analysis plans are signed before unblinding; datasets and programs go in with the filing. This makes it possible to tell when the call was made and to check it but not whether the call was the right one. That instead comes back to two things: standing, which gives the judgment institutional authority, and answerability, which makes someone responsible for explaining it. ICH E9's requirement of a named trial statistician is one instance of that \parencite{ichE9StatisticalPrinciples1998}. 

Standing is what someone gets from where they sit and the profession behind them. This makes standing structural rather than a matter of character, which a regulator or a referee cannot observe anyway. A clinician doing a post-randomization subgroup analysis and declaring the effect causal would be making a mistake; if I did the same, I would be failing at my job. Because the obligation to uphold those standards comes from the profession, professional standards have to outrank any one employer in technical matters \parencite{demingPrinciplesProfessional1965}. This also explains why statistical functions are often placed apart from the teams whose claims they support. 

Standing alone is not enough. Answerability means owing an account of what the data support and of the choices that determined it, to someone entitled to ask for it and to reject it. It is related to accountability, which is answerability plus enforcement \parencite{schedlerConceptualizingAccountability1999}. Organizations are held accountable all the time, but whether anyone in them is answerable is a different question. Knowing that a rejected account will reflect badly on you changes how you make a choice \parencite{lernerAccountingEffectsAccountability1999}. Importantly, this includes the choices no procedure anticipated. We are answerable when we sign an analysis plan, put our name on a paper, or tell a colleague their analysis looks fine. That last one is often sought as an official blessing of decisions already made rather than a review \parencite{brossRoleStatisticianScientist1974}. However, what travels is the assurance that \enquote{I had the statistician look it over}, which is why our obligation is the same. What we owe is an account of which choices were made and why; not \enquote{the software said so}. 

I have defined statisticians here by where they sit and what they owe, not by what they know. Expertise is of course inseparable from standing. The point is that those at the receiving end of a claim cannot verify it, so what reaches them is the position and the obligation.

\section{What Does AI Change?}
\label{sec:ai-change}

AI or no AI, an organization can have a statistical function where every analysis is done fast and flawlessly, yet still have nobody in particular answering for what it ends up claiming. \textcite{schmitzMoralAgencyFramework2025} call this an \enquote{ethics sink}. 

AI can save time on routine statistical work \parencite{doblerChatGPTToolBiostatisticians2025}. Its effect on our profession is a different question, one that has mostly been answered in terms of capability. \textcite{harbronWillPharmaceuticalIndustry2026} lists where pharmaceutical statisticians still add unique value in an AI world: understanding data, understanding uncertainty, rigor, and understanding drug development. \textcite{hoerlFutureStatisticsAI2026} does something similar outside pharma and settles on data quality. I do not disagree with any of it. My reservation is that such lists get shorter with each model release. 

Software has made claim generation cheaper for decades, but AI has lowered the expertise needed to produce something that looks disciplined. A carefully qualified analysis used to require a lot of effort, and effort is a heuristic for quality \parencite{krugerEffortHeuristic2004}. As qualification becomes cheaper, it tells us less about the work behind it. The risk is that we end up discounting even the analyses that deserved better. Boilerplate hedging predates AI, of course, and I have written my share of \enquote{unmeasured confounding cannot be ruled out}. But even with all the outputs in front of me, a cautious tone is hard to read as a sign of quiet deliberation when the analyses underneath took four seconds.

Statisticians are familiar with statistical form detaching from statistical judgment. The null ritual is one example \parencite{gigerenzerMindlessStatistics2004}: a mechanical procedure that did not invite judgment, taught and practiced as statistics itself. It still required people who could owe an account for how it was used.

A model can supply an account, but it cannot owe one. Nothing a model says binds it to a position it may later have to defend. It will not be in the room when the FDA questions come back, and it loses nothing when people remember its mistakes. There is also no institutional weight behind its objections. Say the model gets the statistics right and flags the issues a good statistician would flag. The user may not like the answer anyway. The ideal professional described by \textcite{demingPrinciplesProfessional1965} takes orders in technical matters from the standards of the profession, never from whoever is asking. A model takes orders from the prompt \parencite{jankCanAIReplace2026}, and the user can simply ask it again. I can also be overruled, but not by asking me again.

Putting a statistician at the end of a conveyor belt to sign off on analyses is also not answerability. Take the caveat \enquote{the subgroup analysis should be interpreted with care}. This is not an objective fact about the world. It is a professional judgment based on the subgroups not having been prespecified, there being eleven of them, and the interaction test being underpowered. To be answerable for the caveat, you have to rerun enough of the analysis to reconstruct the judgment behind it which uses up the time AI was supposed to save. If you sign off without reconstructing, it is what \textcite{elishMoralCrumpleZones2019} called a \enquote{moral crumple zone}, where the human formally kept in the loop carries responsibility for decisions they had no capacity to make differently. For answerability to mean anything, it has to be possible to choose otherwise \parencite{schmitzMoralAgencyFramework2025}.

Is answerability just what we ask for while the technology is still unreliable? Suppose a model is correct 99\% of the time. At that point it looks more like an instrument, and nobody asks a calibrated instrument to answer for its readings---we validate it for an intended use and then rely on it. Someone still has to decide what that intended use is. A 99\% accuracy figure describes performance over the tested cases. Deciding that a study is ordinary enough to rely on the model is itself a claim, and it falls under statistical practice. The statistician's responsibility then shifts from checking individual answers to establishing what warrants relying on the model. 

That still fits the definition of statistical practice used here, conveniently so perhaps. Where does it start to break down? \textcite{harbronWillPharmaceuticalIndustry2026} runs a thought experiment: a sponsor sets AI agents to take clinical trial data through analysis and reporting and submit it; the regulator sets its own agents to review the documents, send questions back, and decide. He rejects it because of the \enquote{near certainty of mistakes and hallucinations}. Suppose this problem is solved. The drug is approved. No statistical practice is called for at the level of the individual approval, because the agents have taken care of it. Ask later why this population, or why that safety signal was not pursued, and the answer comes back fast and flawless. Nobody owes an account of those choices for this particular approval, because they were settled at a system level. This is mechanical objectivity made complete: wall-to-wall procedure, open to challenge by no one. But a drug approval is not a pipeline output. It is a decision a regulator stands behind, and its authority comes from having been open to refutation by parties with something at stake \parencite[p.~214]{porterTrustNumbersPursuit1995}. Whether such an approval could warrant trust is a separate question, and not one that solving mistakes and hallucinations answers.

\section{Statistical Leadership After AI}
\label{sec:leadership}

Statistical leadership---deciding where discipline is applied and answering for the way it is applied---is practiced by a statistical function across a portfolio and by a statistician within their own work, whether junior or senior. Every claim involves a decision about how much discipline is required to link it back to data. Sometimes the claim is made by a clinician who decides an analysis needs a statistician on it, sometimes by an AI engineer who decides their new system does not. These are acts of statistical leadership, and they are not reserved for statisticians. Answerability still attaches to the decision, but a clinician or an AI engineer is rarely placed to make good on it. If we take our job to be only the analyses we are asked for, the deciding is done by people who owe an account they cannot give.

As claims become cheap to generate, statisticians cannot be on every single one. So we must seek out the ones that will become consequential, which is harder than finding the ones that already are. When the primary analysis of an important study is discussed, everyone knows what is at stake and the claim comes to us. Most claims do not come labeled like that, and many do not come to us at all. We still need domain knowledge to know which conversations will produce claims that matter and enough informal standing to be in them, whether they happen during a project meeting, at a key trigger point, or around the coffee machine. 

The claims that do come to us are not always the ones to prioritize. The primary analysis of a registrational clinical trial is highly consequential but by the time the analysis actually takes place, key choices are already fixed and we can mostly only improve the account of decisions already made. The phase 2 exploratory analysis that later becomes the dose rationale for a whole development program looks less consequential and is still open. Claims that are left unsupervised tend to outgrow their original warrant. That is how a technically excellent statistics function can be unaware of what its organization is currently asserting on the basis of its own analyses.

A new statistical method is a claim about all the cases it will be applied to, made answerable to the scientific community through publication. Methodology development is therefore an investment in an organization's capacity to answer for claims. An AI system that does statistical work is the same kind of claim: that its built-in discipline is adequate for the cases it is designed to handle. This gives statisticians a design role in these systems, not just a validation role. If judgment can be made mechanically, it can be baked in, and someone answers for having baked it in across every output. If expert judgment remains necessary for each output, the system has to show its work in a form that allows the statistician to reconstruct the judgment and answer for it. Statisticians have been making judgment mechanical for a long time, and there is no exemption for us now. Where judgment can be made mechanical, building the thing that shrinks our role is part of the job.

Return to the Tuesday afternoon: what do I tell my colleague? I can offer an informal view on what they bring, but I cannot answer for it if I cannot reconstruct the judgment behind it. And I have to say which one I am offering, otherwise an informal view can easily be taken as endorsement. If I sit at the end of the conveyor belt like this, I become the \enquote{bad cop} of \textcite{hoerlMovingStatisticsProfession2010}. The alternative is to be in the conversation earlier because ensuring answerability is as much about scoping as about saying no. \textcite{demingPrinciplesProfessional1965} recommended agreeing at the planning stage which decisions belong to the statistician versus the domain expert---whether a dropout pattern is ignorable, say, or what counts as a clinically meaningful difference. That is arguably more important today when the analysis is the easy part. At the level of a statistical function, protecting answerability means backing reasoned refusals and narrower scopes, rather than leaving the cost with the statistician who made the call.

Everything here assumes that our discipline is wanted or at least not unwelcome. In practice, statistical discipline often looks like friction at the point of decision, and nobody wants friction. AI may increasingly offer what people were hoping for in the first place: a very able calculator that does not complain. If we cede the work where our discipline is not wanted, we have to argue for the value of what remains, and much of that only shows up elsewhere and later. The harder sell is that sometimes the friction is the point---and those on the receiving end have to buy that argument as much as we have to make it.

I was a statistical calculator for just a few years. The label has stuck to our profession for rather longer because it has never been entirely unfair. We do pick work off the end of the conveyor belt and we do bless the odd slide deck. AI will make both easier to do and the label harder to shake. We will be tempted to answer with everything else we know. The better answer is older and less flattering: the job is to say what data can and cannot warrant, and to be there when someone asks why.

\section*{Disclosure Statement}
The author is an employee of Novo Nordisk A/S. The views expressed in this article are those of the author and do not necessarily represent the views of Novo Nordisk A/S. 

\printbibliography
 
\end{document}